\documentclass[twocolumn,showpacs,preprintnumbers,amsmath,amssymb,aps,pra]{revtex4}
\usepackage{graphicx,graphics}
\usepackage{dcolumn}% Align table columns on decimal point
\usepackage{bm}% bold math
\usepackage{psfrag}

\begin{document}
\title{Repulsively-bound exciton-biexciton states in high-spin fermions in optical lattices}
\author{A. Arg\"uelles}
\author{L. Santos}
\affiliation{Institut f\"ur Theoretische Physik, Leibniz Universit\"at Hannover, Appelstr. 2 D-30167, Hannover, Germany}
\begin{abstract}
We show that the interplay between spin-changing collisions and quadratic Zeeman coupling 
provides a novel mechanism for the formation of repulsively bound composites 
in high-spin fermions, which we illustrate by considering spin flips in an initially 
polarized hard-core 1D Mott insulator of spin-$3/2$ fermions. 
We show that after the flips the 
dynamics is characterized by the creation of two types of exciton-biexciton composites. 
We analyze the conditions for the existence of these bound states, and discuss their intriguing properties. In particular we show that the effective mass and stability 
of the composites depends non-trivially on spin-changing collisions, on the quadratic Zeeman effect and on the initial exciton localization. Finally, we show  
that the composites may remain stable against inelastic collisions, 
opening the possibility of novel quantum composite phases.
\end{abstract}

%\date{\today}

\maketitle

%%%%%%%%%%%%%%%%%%%%%%%%%

% INTRODUCTION

\section{Introduction}
\label{sec:Introduction}

Ultra cold gases in optical lattices offer an extraordinary controllable environment 
for the analysis of many-body phenomena, as highlighted by the realization of the 
Mott insulator to superfluid phase transition in Bose gases~\cite{Greiner2002}, and 
more recently by the observation of the Mott insulator to metal transition in two component 
Fermi systems~\cite{Jordens2008,Schneider2008}. In addition to a wealth of possible 
quantum phases~\cite{Lewenstein2006,Bloch2008}, and due to its discrete nature, 
optical lattices offer as well novel possibilities for composite formation, including 
local on-site dimer (or $n$-mer) formation in attractive gases~\cite{Rapp2007} and  
fermionic composites in Bose-Fermi mixtures~\cite{Lewenstein2004}. An even more striking 
effect of the lattice discreteness is given by the recent observation of 
repulsively bound atomic pairs (doublons) at a given lattice site~\cite{Winkler2006,Strohmeier2010}, 
which are dynamically stable in the absence of dissipation. Note that contrary to standard 
condensed-matter systems, where e.g. interactions with phonons would lead to a rapid 
dissipation of the dimers, optical lattices provide to a large extent a dissipation-free environment. Indeed experiments on optical lattices provide an extraordinary scenario 
for the study of the intriguing physics of metastable bound states and other far-of-equilibrium phenomena~\cite{Bloch2008}.

Spinor gases, formed by atoms with various available Zeeman substates, present 
a very rich physics due to the interplay between internal and external degrees of freedom~\cite{Ho1998}. Interestingly, interatomic interactions lead to spin-changing processes in which population is transferred between Zeeman sublevels. The corresponding spinor dynamics has attracted a large interest, mostly in the realm of spinor Bose-Einstein condensates~\cite{Schmaljohann2004,Sadler2006,Klempt2010}. Spinor lattice gases 
offer fascinating novel physics, most relevantly in what concerns quantum magnetism, 
including antiferromagnetic order in spin-$1/2$ fermions~\cite{Werner2005} and 
even more intriguing phases for higher spins~\cite{Honerkamp2004,Wu2005,Azaria2005,Gorshkov2010,Cazalilla2009,Jordens2010}. 

\begin{figure}%[ht]
\includegraphics[width=0.45\textwidth]{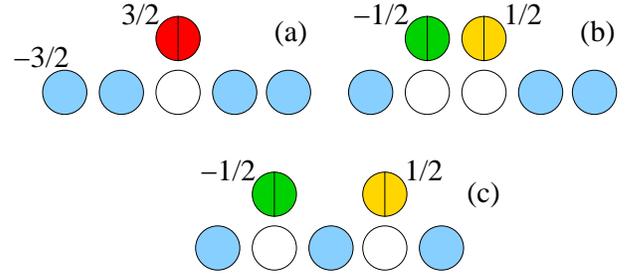}
\vspace*{0.2cm}
\caption{(Color online) (a) Exciton of a $3/2$ particle and a $-3/2$ hole; (b) Biexciton of
by two $m=\pm 1/2$ particles and two $m=-3/2$ holes; (c) Broken biexciton due to $c_2$ processes (see text).}
\vspace*{-0.5cm}
\label{fig:1}
\end{figure}

Repulsively on-site bound pairs in spin-$1/2$ fermions are formed since 
the large interaction energy of the doublon cannot be accommodated by 
the maximal kinetic energy for two atoms in the lowest band
(proportional to the hopping energy)~\cite{Winkler2006,Petrosyan2007,Alzetto2010,Wang2010}. 
In this paper we show that a completely different 
mechanism, based on the interplay between spin-changing and quadratic Zeeman effect (QZE), 
may sustained novel types of composites for repulsive high-spin lattice fermions. 
We illustrate this physics for the case of hard core 1D spin-$3/2$ fermions in a Mott insulator  
initially polarized in $m=-3/2$. We show that spin-flips into $m=3/2$ may lead to 
two types of composites formed by an exciton ($m=3/2$ particle and $m=-3/2$ hole) and an antisymmetric biexciton ($m=\pm 1/2$ particles and two $m=-3/2$ holes) (Fig.~\ref{fig:1}). 
We show that the dynamics and stability of the exciton-biexciton composites exhibits a non-trivial dependence 
with spin-changing, QZE and center-of-mass momentum. Finally, we show that inelastic composite-composite interactions may be very inefficient, opening 
exciting possibilities for many-body composite gases.

%%%%%%%%%%%%%%%%%%%%%%%%%%
% FIGURE 1
%%%%%%%%%%%%%%%%%%%%%%%%%%

The structure of the paper is as follows. In Sec.~\ref{sec:Model} we introduce the spin-$3/2$ system under consideration as well as the possible exciton, and biexciton excitations occurring after an individual 
spin flip. Sec.~\ref{sec:Bound} introduces the idea of repulsively bound exciton-biexciton pairs, with a particular emphasis in their band-like dispersion and the conditions for their 
existence. Sec.~\ref{sec:Dynamics} is devoted to the intriguing dynamics of the repulsively bound pairs, and its dependence on the spatial delocalization of the initial spin-flip excitations. In Sec.~\ref{sec:Multiple} we analyze the case of multiple spin-flip excitations. Finally, in Sec.~\ref{sec:Conclusions} we summarize our conclusions.

%%%%%%%%%%%%%%%%%%%%%%%%%%%%%%%%%%%%%%%%%%%%%%%%%%%%%%%
% 1D SPIN-3/2 FERMIONS IN THE HARD-CORE MOTT REGIME

\section{Model}
\label{sec:Model}

\subsection{Hamiltonian}
\label{subsec:Hamiltonian}

We consider spin-$3/2$ fermions ($m=\pm 3/2,\pm 1/2$) in a  
deep 1D lattice, such that at low filling only the lowest band is relevant. In this regime, the physics  
is given by the interplay between inter-site hopping (characterized by the 
constant $t$), QZE (given by a constant $q$, which may be externally controllable),  
and $s$-wave collisions. The latter may occur in two different channels with total spin $F=0$ and $2$, 
characterized by the coupling constants $g_F=4\pi\hbar^2 a_F/M$, with $a_F$ the scattering length and $M$ the atomic mass. Although typically $a_0$ 
and $a_2$ are similar, they may be varied by microwave dressing~\cite{Papoular2009} or optical Feshbach resonances~\cite{Fedichev1996}. 
Below we use $G=(g_0+g_2)/2$, and $g=(g_2-g_0)/(g_2+g_0)$. 
The $s$-wave interactions preserve magnetization (and hence the linear Zeeman effect is irrelevant), 
but may induce spin-changing collisions (characterized by $g$), which redistribute the populations at different components.
For $G\gg t$ we may consider maximally one fermion per site (hard-core case). 
For a sufficiently large chemical potential the system enters 
into the Mott insulator with one fermion per site. The spin physics within the 
Mott insulator is given by an effective Hamiltonian:
\begin{equation}
\hat H=\sum_{i,m}q m^2 \hat n_{m,i}+\sum_{\langle i,j\rangle}\hat H_{i,j},
\end{equation} 
where $\langle \cdots \rangle$ denotes nearest neighbors and 
\begin{eqnarray}
&&\!\!\!\hat H_{i,j}=c_2\!\!\sum_{|m| \neq |m'|}\left (
\hat n_{m,i}\hat n_{m',j}-\hat \psi_{m,i}^\dag\hat \psi_{m',j}^\dag\hat \psi_{m,j}\hat \psi_{m',i}
\right ) \nonumber \\
&&\!\!\!+\!\!\!\sum_{|m|=1/2}^{3/2}\!\!c_{|m|}\!\left (
\hat n_{m,i}\hat n_{-m,j}-\hat \psi_{m,i}^\dag\hat \psi_{-m,j}^\dag\hat \psi_{m,j}\hat \psi_{-m,i}
\right )\nonumber \\
&&+c_{sc}\left [
\left (\hat\psi_{-1/2,i}^\dag\hat\psi_{1/2,j}^\dag-\hat\psi_{1/2,i}^\dag\hat\psi_{-1/2,j}^\dag \right )\right\delimiter 0 \nonumber \\
&&\left\delimiter 0 \left (\hat\psi_{3/2,j}\hat\psi_{-3/2,i}-\hat\psi_{-3/2,j}\hat\psi_{3/2,i} \right )
+H.c. \right ],
\label{eq:Hij}
\end{eqnarray}
describes the effects of the interatomic interactions, with
\begin{eqnarray}
&&c_{3/2}\equiv -2t^2 \left (
\frac{\cos^2\phi}{9q/2+\lambda_+}+\frac{\sin^2\phi}{9q/2+\lambda_-}
\right ), \\
&&c_{1/2}\equiv -2t^2 \left (
\frac{\cos^2\phi}{q/2+\lambda_-}+\frac{\sin^2\phi}{q/2+\lambda_+}
\right ), \\
&&c_{\mathrm{sc}}\equiv t^2 \sin 2\phi \sum_{\beta=\pm} \eta_\beta
\frac{5q/2+\lambda_\beta}{\prod_{{\bar m}=1/2}^{3/2}(2q{\bar m}^2+\lambda_\beta)}, \\
&& c_2=-t^2/(g+G), \label{eq:c2}
\end{eqnarray}
where $\eta_\pm=\pm 1$, 
$\lambda_\pm\equiv G-5q/2\pm \left [ 4q^2+g^2G^2 \right ]^{1/2}$, and 
$\tan\phi=\left([4q^2+g^2G^2]^{1/2}+2q\right )/gG$
%%%
~\cite{footnote}. The constant $c_{3/2}$ ($c_{1/2}$) characterizes the super-exchange between 
neighboring sites with $\pm 3/2$ ($\pm 1/2$), whereas $c_2$ does the same for sites 
with $m$ and $m'$ such that $|m|\neq |m'|$. Finally, $c_{sc}$ characterizes the super-exchange 
leading to spin-changing from $\pm 1/2$ to $\pm 3/2$ or viceversa.

%%%%%%%%%%%%%%%%%%%%%%%%%%%%%%%%%%%%%%%%%%%%%%%%%%%%%%%

% SINGLE FLIP EXCITATIONS

\subsection{Single-flip excitations}
\label{subsec:Single-flip}

We consider a system initially polarized into $m=-3/2$, 
$|\psi_{BG}\rangle =\prod_i \psi^\dag_{-3/2,i}|vac\rangle$, where $|vac\rangle$ is the vacuum. 
$|\psi_{BG}\rangle$ is stable since collisions preserve magnetization. 
We are interested in excitations of the form  
$|m,j\rangle\equiv\hat\psi_{m\neq -3/2,j}^\dag\hat\psi_{-3/2,j}|\psi_{BG}\rangle$, created 
by spin flips. Note that these excitations 
have actually an {\em exciton-like character}, since they are formed by 
a particle (in $m\neq -3/2$) and a hole (in $m=-3/2$) (Fig.~\ref{fig:1}a). 

A single spin-flip into $m=\pm 1/2$ is not accompanied by spin-changing, 
and hence lead to isolated $\pm 1/2$ excitons which hop with an effective hopping $c_2$. 
Spin-changing collisions lead to a much less trivial dynamics when spins are flipped into $m=3/2$.
After the spin-flip the $3/2$ exciton may hop to the nearest neighbor via 
$c_{3/2}$ super-exchange~(second line in Eq.~\eqref{eq:Hij}). In absence of spin-changing  
($g=0$) the $3/2$ exciton propagates freely with an energy 
$E_{3/2}(k)=2c_{3/2}(1-\cos k)$, where $k$ is the center-of-mass momentum of the $3/2$ exciton.

Spin-changing modifies this simple picture, since the $m=3/2$ spin may 
interact via $c_{sc}$ super-exchange with a neighboring $-3/2$ 
spin leading to two neighboring spins with $m=\pm 1/2$, i.e. a biexciton 
of the form $|m,j;-m, j+1\rangle \equiv \rangle\equiv \psi_{m,j}^\dag\psi_{-m,j+1}^\dag\psi_{-3/2,j+1}\psi_{-3/2,j}|\psi_{BG}\rangle$~(Fig.~\ref{fig:1}b). 
After spin-changing, one of the newly created excitons, say $m=1/2$, may move apart from the neighboring $m=-1/2$ one by a $c_2$ superexchange~(Fig.~\ref{fig:1}c). Further $c_2$ processes bring the 
pair even further away from each other. Since an isolated $\pm 1/2$ spin surrounded by $-3/2$ neighbors 
remains stable against spin-changing, one naively expects all $3/2$ 
excitons to eventually dissolve into independent $\pm 1/2$ ones, which would freely move away 
with a hopping $c_2$. As shown below, contrary to the naive expectation, the $3/2$ excitons may become very 
robust.

%%%%%%%%%%%%%%%%%%%%%%%%%%%

\section{Exciton-biexciton repulsively bound states}
\label{sec:Bound}

\subsection{Band structure}
\label{subsec:Band}

We introduce the Fourier transformed states~\cite{Nguegang2009} 
$|\phi_{j}^{\pm}(k)\rangle\!\!\! \equiv\!\!\! \frac{1}{\sqrt{L}}\sum_{l=1}^L e^{ik(l+j/2)}| \mp 1/2,l; \pm 1/2,l+j\rangle$ and 
$|\psi_{3/2}(k)\rangle\equiv\frac{1}{\sqrt{L}}\sum_{l=1}^L e^{ikl}|3/2,j\rangle$,
where $L$ is the number of sites and $k$ the center of mass quasi-momentum either of the $3/2$ exciton or of the $\pm 1/2$ biexciton. Note that the biexciton states  $|\phi_{j}^{\pm}(k)\rangle$ contain not only 
biexcitons of neighboring $\pm 1/2$ (i.e. $j=1$, as in Fig.~\ref{fig:1}b), but also broken 
biexcitons~(i.e. $j>1$, as in Fig.~\ref{fig:1}c). This is true even for the bound solutions discussed 
before, which may contain as well a contribution of broken biexciton pairs. 

It is particularly useful to introduce the symmetric and antisymmetric states $|\phi_j^{S,A}(k)\rangle=(|\phi_{j}^+(k)\rangle\pm |\phi_{j}^+(k)\rangle)/\sqrt{2}$.
Note that the Hamiltonian~\eqref{eq:Hij} do not couple symmetric and antisymmetric 
states, and hence the Hamiltonian in $k$-space splits into 
$H=\sum_k (H_S(k)+H_A(k))$. The physics of the symmetric biexcitons is given by 
\begin{eqnarray}
H_S(k)&=&E_{1S} |\phi_1^S(k)\rangle\langle\phi_1^S(k)|
+E_{B}\sum_{j>1}|\phi_j^S(k)\rangle\langle\phi_j^S(k)| \nonumber \\
&-&\Omega_2(k)\sum_{j\geq 1}(|\phi_j^S(k)\rangle\langle\phi_{j+1}^S(k)| + H.c.),
\label{eq:HS}
\end{eqnarray}
with $E_{1S}=2c_2-4q$, $E_B=4c_2-4q$ and $\Omega_2(k)=2c_2\cos(k/2)$, whereas 
the coupling between $3/2$ excitons and 
antisymmetric $\pm 1/2$ biexcitons is given by
\begin{eqnarray}
H_A(k)&=&E_{1A} |\phi_1^A(k)\rangle\langle\phi_1^A(k)|+
E_B \sum_{j>1} |\phi_j^A(k)\rangle\langle\phi_j^A(k)| \nonumber \\
&+& E_{3/2}(k)|\psi_{3/2}(k)\rangle\langle\psi_{3/2}(k)|\nonumber \\
&+&\Omega_{sc}(k) (i|\psi_{3/2}(k)\rangle\langle\phi_1^A(k)|+H.c.)
 \nonumber \\
&-& \Omega_2(k)\sum_{j\geq 1}(|\phi_j^A(k)\rangle\langle\phi_{j+1}^A(k)| + H.c.),
\label{eq:HA}
\end{eqnarray}
where $E_{1A}=2c_2+2c_{1/2}-4q$, $E_{3/2}(k)=2c_{3/2}(1-\cos k)$, 
and $\Omega_{sc}(k)=\sqrt{8}c_{sc}\sin (k/2)$. 

%%%%%%%%%%%%%%%%%%%%%%%%%%
% FIGURE 2
\begin{figure}%[ht]
\hspace*{-0.3cm}
\includegraphics[width=0.4\textwidth]{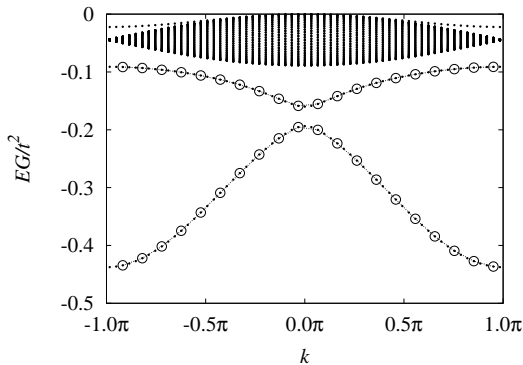}
\vspace*{-0.2cm}
\caption{Band spectrum for single-flip excitations as a function the quasimomentum $k$, 
for $\tilde q=-4$ and $g=0.8$. The circles indicate the results of Eq.~\eqref{eq:trasc}.}
\label{fig:2}
\vspace*{-0.5cm}
\end{figure}
%%%%%%%%%%%%%%%%%%%%%%%%%%

Figure~\ref{fig:2} shows a typical spectrum (which resembles that of 
self-bound repulsive pairs of spin-$1/2$ fermions~\cite{Winkler2006}).  
Unbound $\pm 1/2$ biexcitons form a continuous band, due to their free 
relative motion with an effective hopping $\Omega_2(k)$, which leads to a dispersion $E(k,k_r)=E_B + 2\Omega_2(k)\cos {k_r}$, where $k_r$ is the relative momentum of the pair. 
We observe also the appearance of up to three bound states~(resembling the case of 
repulsive spin-$1/2$ fermions with nearest neighbor interactions~\cite{Nguegang2009}). 
One of the bound states (upper state in Fig.~\ref{fig:2}), with energy $E(k)=E_B-2c_2(1+\cos^2(k/2))$, belongs to the symmetric manifold, being hence decoupled from the $3/2$ exciton, and thus we do 
not consider it further.

%%%%%%%%%%%%%%%%%%%%%%%%%%%%%%%%%%%%%%%%%%%%%%%%%%%%%%%

\subsection{Pairing of excitons and antisymmetric biexcitons}
\label{subsec:Pairing}

% ANTISYMMETRIC-3/2 BOUND STATES

\begin{figure}%[ht]
\vspace*{-0.7cm}
\includegraphics[width=0.45\textwidth]{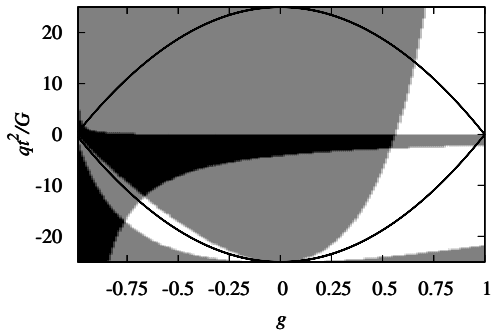}
\vspace*{-0.5cm}
\caption{Regions with $0$ (black), $1$ (grey) and $2$ stable exciton-biexciton 
composites for $k=0$, as a function of $g$ and $qt^2/G$ for $t=0.1G$.}
\label{fig:3}
 \vspace*{-0.5cm}
\end{figure}

The other two bound states result from the coupling 
of antisymmetric $\pm 1/2$ biexcitons and $3/2$ excitons. 
In order to understand these states, we introduce the ansatz 
\begin{eqnarray}
|\psi(k)\rangle&=&\cos\varphi\sum_{j\geq 1} (-1)^{j-1}e^{-(j-1)/\gamma} |\phi_j^A(k)\rangle
\nonumber \\
&+&i\sin\varphi|\psi_{3/2}(k)\rangle, 
\label{eq:wavef}
\end{eqnarray}
where $\gamma$ is the localization length. 
Imposing $E|\psi(k)\rangle=H_A(k)|\psi(k)\rangle$ results in a transcendent equation 
for $\gamma$:
\begin{eqnarray}
&&E_B+2\Omega_2 \cosh(1/\gamma)=  \nonumber \\
&&\frac{1}{2}\left [ E_{1A}+\Omega_2 e^{1/\gamma}+E_{3/2} \right ]
\nonumber \\
&&\pm \frac{1}{2}\sqrt{
\left [ E_{1A}+\Omega_2 e^{1/\gamma}-E_{3/2} \right ]^2+4\Omega_{sc}^2
}, 
\label{eq:trasc}
\end{eqnarray}
where the relation between energy and $\gamma$ is given by 
\begin{equation}
E(k)-E_B = 2 \Omega_2(k) \cosh (1/\gamma(k)).
\label{eq:E-EB}
\end{equation}
In Fig.~\ref{fig:2} we indicate the results obtained from Eqs.~\eqref{eq:trasc} and~\eqref{eq:E-EB}, which 
are in excellent agreement to those directly obtained from the diagonalization of~\eqref{eq:HA}. 
Eq.~\eqref{eq:E-EB} shows that the separation between the bound state and the lowest 
boundary $E_B+2\Omega_2(k)$ of the sea of unbound $\pm 1/2$ pairs is crucial for its localization, and hence 
while the lower bound state is tightly bound, the upper one is much more loose.
In particular, the upper bound state unbinds ($\gamma$ diverges) if its energy approaches 
$E_B+2\Omega_2$. Hence the existence of the bound exciton-biexciton states depends non-trivially 
on QZE, spin-changing collisions, and exciton momentum. 
Fig.~\ref{fig:3} shows those values of $g$ and $q$ for which one or two bound states 
are expected for $k=0$.

%%%%%%%%%%%%%%%%%%%%%%%%%%
% FIGURE 3
%%%%%%%%%%%%%%%%%%%%%%%%%%

\section{Dynamical properties of the exciton-biexciton bound pairs}
\label{sec:Dynamics}

The mixing angle $\varphi$ fulfills
\begin{equation}
\tan\varphi(k)\!=\!\frac{1}{\Omega_{sc}(k)} \left ( E(k)-E_{1A}-\Omega_2(k)e^{-1/\gamma(k)} \right ). 
\end{equation}
The contribution of the $3/2$ exciton to the bound states ($\sin^2\varphi$) is particularly 
important to understand the dynamics following a spin-flip into $m=3/2$. In particular, at $k=0$ (when the center of mass of the spin-flip 
excitation is maximally delocalized), $\Omega_{sc}=0$ and hence the upper bound state is a pure 
antisymmetric $\pm 1/2$ biexciton with $E=E_{1A}+4c_2^2/E_{1A}$ and  
$\gamma=1/\ln (E_{1A}/2c_2)$, whereas the lower bound state is a pure $m=3/2$ 
exciton with energy $E_{3/2}(0)$. On the contrary, when the center of mass of the spin-flip excitation is strongly localized 
there is a strong coupling between the exciton and the antisymmetric biexciton which leads to a strongly distorted dynamics, as shown below.

%%%%%%%%%%%%%%%%%%%%%%%%%%
% FIGURE 4
\begin{figure}%[ht]
\vspace*{-0.4cm}
\includegraphics[width=0.35\textwidth]{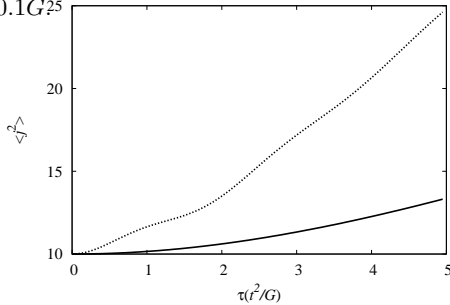}
\vspace*{-0.2cm}
\caption{Time evolution of the width $\langle j^2 \rangle$ of the $3/2$ 
exciton wavepacket for $t=0.1G$, $g=0.8$ and $qt^2/G=-4$ with~(dotted) and without~(solid) spin-changing collisions, for an initial Gaussian wavepacket with a width of $10$ sites.}
\label{fig:4}
% \vspace*{-0.5cm}
\end{figure}
%%%%%%%%%%%%%%%%%%%%%%%%%%

%%%%%%%%%%%%%%%%%%%%%%%%%%%%%%%%%%%%%%%%%%%%%%%%%%%%%%%

% EXCITON WAVEPACKET DYNAMICS

If the initial spin-flip is delocalized in a spatial region 
much larger than the inter-site spacing the momentum distribution 
is strongly peaked at $k=0$. Hence after the spin-flip 
there is no significant transfer into $\pm 1/2$ biexcitons, since 
the lower bound state is basically a $3/2$ exciton. 
Without spin-changing the exciton tunnels with hopping $c_{3/2}$, 
and hence at $k=0$ the effective exciton mass is $m_{*0}=1/2c_{3/2}$.
In spite of the absence of any significant biexciton 
admixture, spin-changing strongly modifies the effective exciton mass into: 
\begin{equation}
\frac{1}{m_*}\simeq 2 c_{3/2}-\frac{4c_{sc}^2}{E_{1A}}.
\end{equation}
Note that $m_*$ differs significantly from $m_{*0}$ (it may even invert its sign), radically 
modifying the exciton wavepacket dynamics, as illustrated in Fig.~\ref{fig:4}. 

\begin{figure}%[ht]
\vspace*{-0.4cm}
\includegraphics[width=0.45\textwidth]{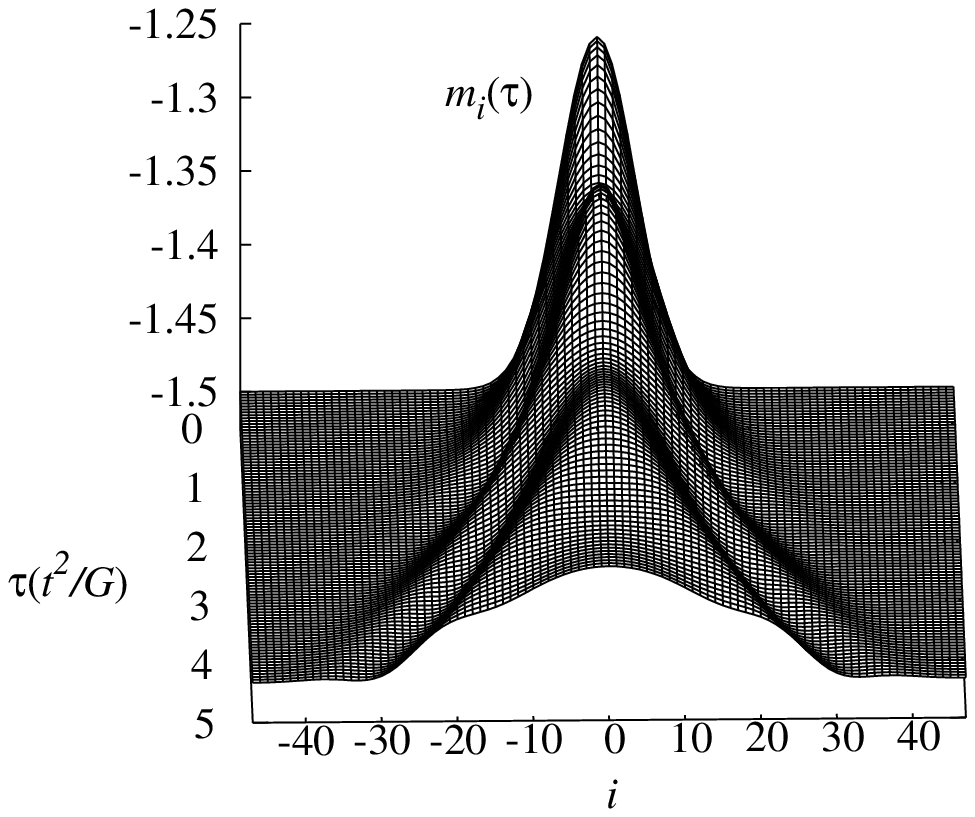}
\includegraphics[width=0.45\textwidth]{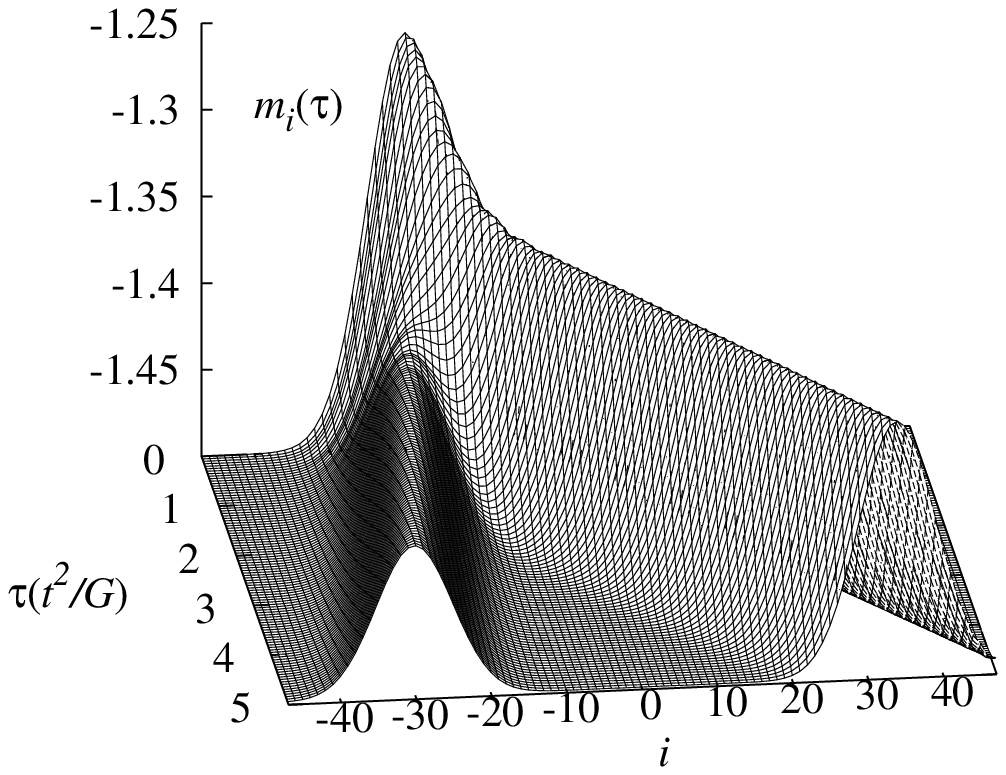}
\includegraphics[width=0.45\textwidth]{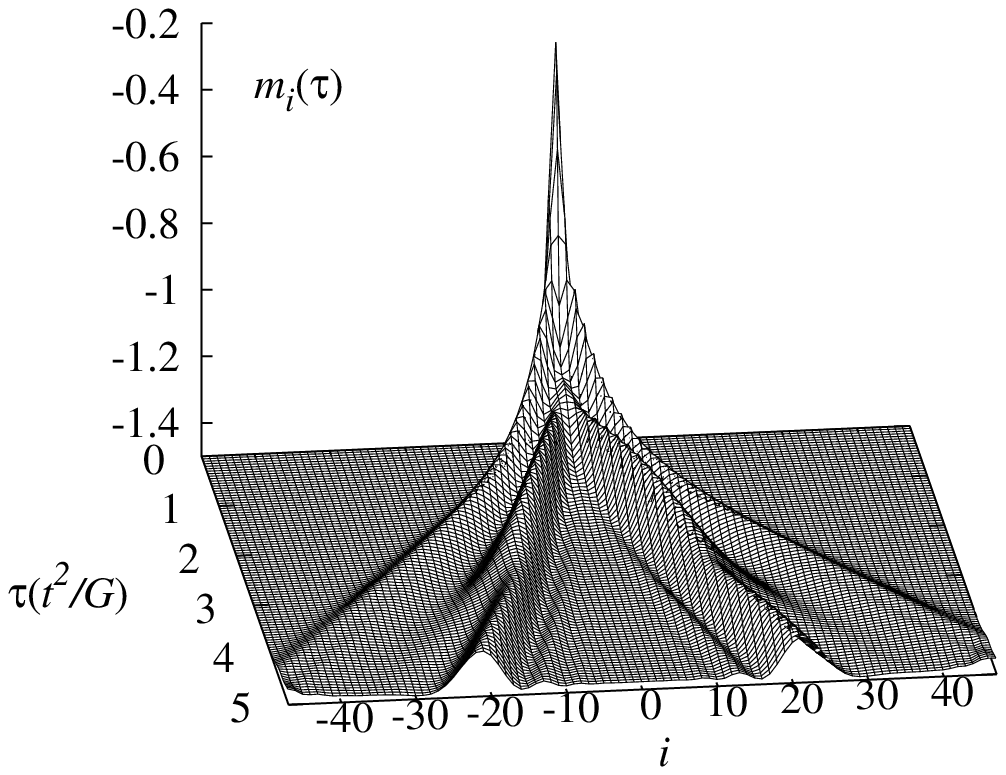}
\vspace*{-0.2cm}
\caption{Time evolution of the magnetization $m_i=\sum_m m \langle \hat n_{m,i} \rangle$ for an initial 
Gaussian wavepacket for the flip excitation with width of $10$ sites and central momentum $k_0=0$ (top), and 
$k_0=\pi/2$ (middle), and for a localized initial flip with width of $1$ site and $k_0=0$ (bottom)}
\label{fig:5}
% \vspace*{-0.5cm}
\end{figure}

Note that if the created initial exciton has a sharply defined but finite center-of-mass momentum $k_0$, 
then the $3/2$ exciton is not any more a single bound state but a linear superposition of both bound states 
(and possibly a slight contribution of broken biexcitons). Since the bound states have different 
group velocities of opposite sign, the dynamics after the spin-flip is hence 
characterized by the appearance of two wavepackets moving in opposite directions formed 
by the two different types of exciton-biexciton composites (Fig.~\ref{fig:5}, middle, which should be compared 
with the single wavepacket spreading for $k_0=0$ in Fig.~\ref{fig:5}, top).

%%%%%%%%%%%%%%%%%%%%%%%%%%
% FIGURE 5
%%%%%%%%%%%%%%%%%%%%%%%%%%

The situation is markedly different if the initial spin-flip is strongly localized, where all $k$ in the Brillouin zone contribute. As a consequence, the initial $3/2$ excitation results 
in an admixture of both bound states, which result, as shown in Fig.~\ref{fig:5} (bottom), to a completely different wavepacket dynamics as that for initially extended spin-flips~(Fig.~\ref{fig:5}, top).
Note that, in addition, a localized initial spin-flip may become considerably more unstable against 
unbinding, especially if the upper antisymmetric-$3/2$ bound state unbinds for 
a given range of $k$ values. In that case a significant part of the initial exciton excitation 
would dissolve into unbound biexciton pairs.

%%%%%%%%%%%%%%%%%%%%%%%%%%%%%%%%%%%%%%%%%%%%%%%%%%%%%%%

% MANY BODY PHYSICS

\section{Multiple spin-flip excitations}
\label{sec:Multiple}

Up to this point we have considered the dynamics after single spin-flips.
Although single spin flips may be created using state of the art techniques for single-site 
addressing~\cite{Bakr2009,Sherson2010}, in typical experiments more than one spin-flip will be induced. 
As a result more than one composite will be simultaneously created, hence 
opening the possibility of inelastic composite-composite collisions which may induce 
a decay channel for the bound composites into broken biexciton pairs.
The various initial spin-flip excitations result in a complicated many-body 
non-equilibrium dynamics for the created bound composites, which we have studied by means of 
time-dependent Matrix-Product-State calculations~\cite{Verstraete2004} 
for up to $5$ spin-flips delocalized in an initial extension of $10$ sites (for a total of $30$ sites).

%%%%%%%%%%%%%%%%%%%%%%%%%%
% FIGURE 6
\begin{figure}[!ht]
%\vspace*{-0.7cm}
\includegraphics[width=0.45\textwidth]{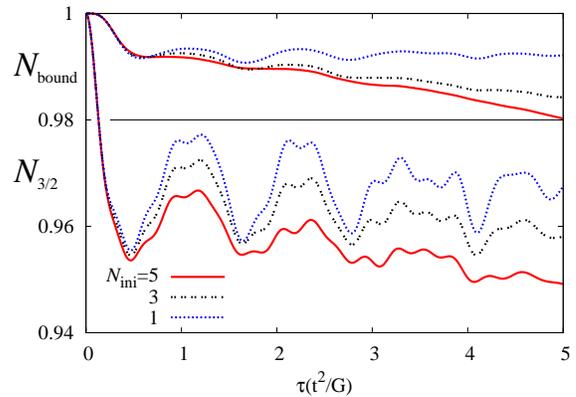}
%\vspace*{-0.3cm}
\caption{(Color online) Time evolution of the bound population (top) and of the population of 
$3/2$ excitons (bottom) for $t=0.1G$, 
$g=0.6$ and $qt^2/G=-4$, for the case of $1$ (dotted), $3$ (dashed) and $5$ (solid) 
spin flips delocalized in an initial square wavepacket $10$ sites wide.}
\label{fig:6}
\vspace*{-0.5cm}
\end{figure}
%%%%%%%%%%%%%%%%%%%%%%%%%%

Figure~\ref{fig:6} shows our results for the case of $qG/t^2=-4$ and $g=0.6$. For this particular 
case the bound states are well below the sea of broken pairs. Hence, as discussed above, it is expected 
that single spin-flip excitations do not lead to a significant population of unbound solutions. 
Two main features can be observed in Fig.~\ref{fig:6}. On one side, there are marked Rabi like 
oscillations of the $3/2$-exciton population due to the presence of the two bound states. The oscillation amplitude is small since, as discussed in the previous section, the lowest bound state is to a large extent the $3/2$-exciton due to the initial delocalization of the spin-flip. Note that the Rabi oscillations are however damped due to dephasing, since different momentum components close to $k_0=0$ have a slightly different Rabi frequency. Note that a more localized initial spin-flip would result in a stronger frequency beating, 
due to the participation of momenta over all the Brillouin zone. Note also that the presence of additional spin-flips leads to an even stronger damping of the Rabi oscillations.

On the other side, note that composite-composite collisions basically leave 
unaffected the total population in the bound states, which for this particular case basically reduces to the 
population of $3/2$ excitons~(Fig.~\ref{fig:1}a) and unbroken $\pm 1/2$ biexcitons~(Fig.~\ref{fig:1}b). Hence, deeply bound states 
may be created, which do not undergo significant inelastic unbinding due to inelastic collisions. 
Thus, multiple spin-flips lead to an out-of-equilibrium (but highly metastable) many-body state, 
which may understood as a basically stable gas of bound composites, opening interesting perspectives 
for quantum composite gases. For the particular case 
considered in Fig.~\ref{fig:6} we expect a quantum gas of almost pure $3/2$-excitons with "biexciton dressed" 
dynamics. Note that although the $3/2$-excitons present an effective attractive nearest neighbor interaction ($-2c_{3/2}$), this interaction becomes irrelevant at low momenta $k\rightarrow 0$ 
compared to the infinite on-site repulsion due to Pauli exclusion. Hence the low $k$ 
properties of 1D $3/2$ excitons will be as those of a Tonks-Girardeau gas.

%%%%%%%%%%%%%%%%%%%%%%%%%%%%%%%%%%%%%%%%%%%%%%%%%%%%%%%

% CONCLUSIONS

\section{Conclusions}
\label{sec:Conclusions}
In summary, spin flips in a polarized repulsive high-spin Fermi gas may lead to the 
formation of novel types of bound composites, which we illustrated for the case of 1D 
spin-$3/2$ hard-core fermions in the Mott phase. In that case the composites are formed by a $3/2$ exciton-like excitation and an antisymmetric $\pm 1/2$ biexciton-like one. Intriguing dynamics and stability properties of the composites result from a non trivial interplay between spin-changing, QZE and exciton momentum. The stability of the exciton gas against inelastic interactions opens exciting possibilities for the creation of 
intricate novel quantum composite phases.  

\acknowledgments
We thank the DFG for support (Center of Excellence QUEST).


\begin{thebibliography}{99}
\bibitem{footnote} A divergence occurs when 
$q=\pm G(g^2-1)/4$, where perturbation theory breaks down, since the 
Mott insulator energy equals that with a particle-hole excitation.
We avoid this divergent case.
\end{thebibliography}
\end{document}